\documentstyle[prb,aps,epsf]{revtex}
\begin{document}
\draft
\twocolumn[
\begin{center}
{\large\bf Nonlocal conductivity in the vortex-liquid regime}\\
\vspace{3ex}
T. Blum and M. A. Moore \\
{\em Department of Physics, University of Manchester,
Manchester, M13 9PL, United Kingdom.}\\
(\today )
\end{center}
\widetext
\begin{abstract}
\leftskip 54.8pt
\rightskip 54.8pt
We investigate the nonlocal conductivity calculated from the
time-dependent Ginzburg-Landau equation.
 When fluctuations of the vector potential are negligible, the
high-temperature (Gaussian) and low-temperature (flux-flow) forms of
the uniform conductivity within an $ab$ plane ($\sigma_{yy}$) are
essentially identical.
 We find to what extent the nonlocal conductivity shares this feature.
 The results suggest that for pure samples in these regimes the length scales
of the nonlocal resistivity, $\rho_{yy}(y-y^{\prime},z-z^{\prime})$, remain
short-ranged in the $y$ and $z$ directions in contrast to the assumptions
made by the hydrodynamic modelling of the multiterminal transport
measurements.
 On the other hand, the resistivity is seen to have a long length
scale in the $x$ direction $\rho_{yy}(x-x^{\prime})$.
 The implications for the interpretation of recent experiments are
discussed.
\end{abstract}
\leftskip 54.8pt
\rightskip 54.8pt

\narrowtext
]

The conductivity is nonlocal if the current at a site ${\bf r}$ is
determined not only by ${\bf E}({\bf r})$ the electric field at ${\bf r}$
but also by the fields at other sites ${\bf E}({\bf r}^{\prime})$.
 It is always nonlocal on some microscopic scale, but the term ``nonlocal"
is generally reserved for cases in which the length scale involved is much
longer than some underlying microscopic one.
 A nonlocal conductivity implies a nonlocal resistivity (though the length
scales on which each varies may differ).
 Since the resistivity of type II superconductors will be influenced
by the motion of flux lines, it seems reasonable to expect nonlocal effects.
 In recent experiments on heavily twinned samples of YBCO, Safar
{\it et al.}\cite{Safar} have observed the onset and enhancement of
nonlocal conductivity as the temperature is lowered.

 With experiments on clean samples, samples with columnar defects and
films underway, it would be useful to sort out the basic form of the
conductivity.
 But the vortex-liquid state of a type II superconductor is a problem with
many length scales and identifying which is/are relevant to a particular
measurement is by no means trivial.

In the Ohmic (linear response) regime, nonlocality implies that the (dc)
conductivity $\sigma_{\mu \nu}({\bf r},{\bf r^{\prime}}) \neq \sigma_{\mu
\nu} \delta({\bf r}-{\bf r^{\prime}})$ or in momentum space: $\sigma_{\mu
\nu}({\bf k}) \neq {\rm const.}$
 Huse and Majumdar\cite{HusMaj} explored a ``hydrodynamic"
approach\cite{MarNel1,MarNel2}, in which the small-${\bf k}$ expansion of the
conductivity is truncated as follows:
\begin{equation}
\sigma_{\mu \nu}({\bf k}) \ = \ \sigma_{\mu \nu}(0)
\ + \ S_{\mu \alpha \beta \nu}~k_{\alpha}k_{\beta}.
\label{hydro}
\end{equation}
In the example they work out in detail to explain features of the
multiterminal transport measurements, they include only one $S$:
$S_{yzzy}$.
 It models the ``tilt viscosity" --- which measures the influence of
pancake vortices moving in one $ab$ plane on those moving in another
when they travel with different velocities.\cite{HusMaj}
 The hydrodynamic picture\cite{MarNel1,MarNel2} began by viewing vortices
as the analogues of polymers in a melt, concepts like viscosities naturally
followed.
 Thus this picture would seem most appropriate in the London limit where
the core size and magnetic screening length $\lambda$ are shorter than the
distance between vortices, {\it i.e.} where vortices are most polymer-like.
 It is unclear, however, what fraction, if any, of the vortex-liquid regime
is described by such a hydrodynamic approach.

 Stability requires $\sigma_{\mu \nu}({\bf k})$ to be a positive definite
matrix; for the hydrodynamic form of the conductivity, it implies for
instance that $S_{yzzy}$ and $S_{yyyy}$  be positive.
 When these conditions are met, one finds that the {\it resistivity}
decays with a length scale of order ($\sqrt{S/\sigma}$), which is
in turn the length scale found in the electric-field and current
distributions.\cite{HusMaj}
 Recent calculations\cite{BM,Mouetal} starting from the time-dependent
Ginzburg-Landau (TDGL) equation find that many of the $S$'s do not have
the requisite sign --- at least for clean  samples at high temperatures.
 We study the nonlocal conductivity within the $ab$ plane $\sigma_{yy}
({\bf k})$ in the lowest Landau level (LLL) limit, which is valid not
close to $H_{c2}(T)$ as is sometimes stated in the literature but at lower
temperatures in the vortex-liquid regime.\cite{Ikeda1}
 When fluctuations of the vector potential are negligible, the uniform
conductivity $\sigma_{yy}({\bf 0})$ is known to interpolate smoothly between
its high-temperature (Gaussian) form and its low-temperature (flux-flow)
form.\cite{Ikeda2}
 We extend these results to the nonlocal conductivity and argue that
$S_{yzzy}$ and $S_{yyyy}$ do not change sign as the temperature is lowered,
suggesting that there may be a substantial region where the
hydrodynamic approach does not apply.

Our starting point is the TDGL equation for the superconducting order
parameter $\Psi$ in a constant magnetic field along the $z$ axis in the
Landau gauge:
\begin{eqnarray}
&&\Biggl[{1 \over \Gamma} {\partial \over \partial t}
- {\hbar^2 \over 2m_{ab}}
\left({\partial^2 \over \partial x^2}+
\left({\partial \over \partial y}
-{ie^*Bx \over \hbar}\right)^2\right) \nonumber \\
&&-{\hbar^2 \over 2m_c} {\partial^2 \over \partial z^2}+a  \Biggr]
\Psi({\bf r},t) +b |\Psi({\bf r},t)|^2 \Psi({\bf r},t) =
\eta({\bf r},t),
\label{TDGL}
\end{eqnarray}
where the $\langle \eta ({\bf r},t)\rangle=0$ and has $\delta$-function
correlations
\begin{equation}
\left\langle \eta^*({\bf r},t)~\eta({\bf r^{\prime}},t^{\prime})
\right\rangle = {2 k_BT \over \Gamma}
{}~\delta({\bf r}-{\bf r^{\prime}}) ~\delta(t - t^{\prime}).
\label{noise}
\end{equation}

At high temperatures, the nonlinear term ($\propto b$) in (\ref{TDGL})
can be ignored.
 The ``states" of the linear operator collapse into Landau levels
\begin{equation}
\phi_n({\bf r})= e^{ik_yy+ik_zz} ~u_n(x/\ell - k_y \ell),
\label{state}
\end{equation}
where $u_n(v)$ are the harmonic-oscillator eigenstates and
$\ell=(\hbar/e^*B)^{1/2}$ the magnetic length.
 The corresponding energies are
\begin{equation}
E_n(k_z)=\hbar^2 k_z^2/2m_c + \alpha + n \hbar \omega_0,
\label{energy}
\end{equation}
where $\omega_0=e^*B/m_{ab}$ is the cyclotron frequency and
$\alpha=a+ \hbar \omega_0 /2$ the energy of the LLL.

 Some length scales are already apparent; clearly $\ell$ is an
$ab$-plane length scale related to the distance between flux lines.
 Two energy scales, $\alpha$ and $\hbar \omega_0$, appear in $E_n(k_z)$;
from these one can construct two $c$-axis length scales:
$\xi_c=(\hbar^2/2m_c \alpha)^{1/2}$ and $\ell_c = (\hbar / m_c
\omega_o)^{1/2}=(m_{ab}/m_c)^{1/2} \ell$.
 The former is the mean-field $c$-axis correlation length; the latter
a $c$-axis analogue to the magnetic length.
 The lengths, $\ell$ and $\ell_c$, depend only on the magnetic field; while
$\xi_c$ grows as the temperature ($T$) is lowered.

The Kubo formula can be used to determine the contribution of the
superconducting fluctuations to the dc conductivity (the Aslamazov-Larkin
term):
\begin{eqnarray}
\sigma^{(s)}_{\mu \nu}({\bf k})&=&{1 \over 2 k_B T}
\int d{\bf r^{\prime}} ~d t^{\prime}
{}~{\rm e}^{i{\bf k}\cdot {\bf r^{\prime}} }
\nonumber \\
&& \ \ \ \times ~\langle J^{(s)}_{\mu}({\bf r+r^{\prime}},t+t^{\prime})
{}~J^{(s)}_{\nu}({\bf r}, t) \rangle_c~.
\label{kubo}
\end{eqnarray}
The superconducting currents are given by:
\begin{equation}
J^{(s)}_{\mu}({\bf r},t)={e^* \hbar \over 2 m_{\mu}}\left[\Psi^*\left(
-i {\partial \over \partial r_{\mu}}-e^*A_{\mu}\right)\Psi+{\rm c.c.}
\right].
\label{current}
\end{equation}
Each involves a $\Psi$, a $\Psi^*$ and a (gauge-invariant) derivative, so
when performing the average in (\ref{kubo}), one ``contracts"
$\Psi({\bf r})$ with $\Psi^*({\bf r^{\prime}})$ and likewise
$\Psi^*({\bf r})$ with $\Psi({\bf r^{\prime}})$ resulting in two
correlators.
In diagrammatic form the result is a loop, the first term in Fig. 1(a).

 The Gaussian result for $\sigma_{yy}^{(s)}({\bf 0})$ is
\begin{equation}
\sigma^{(s)}_{yy}({\bf 0}) \ = \ {e^{*2}k_BT m_c^{1/2} \over
2^{5/2} \pi \hbar^3 \Gamma \alpha^{1/2}} \sum_{n=0}^{\infty}
{(-1)^n(n+1) \over (1 +n\omega_0/2\alpha)^{1/2}}.
\label{Gaussian}
\end{equation}
We will not reproduce the full expression for the Gaussian
$\sigma^{(s)}_{\mu \nu}({\bf k})$ here as they appear in Ref. [5].
 Terms in the small-${\bf k}$ expansion are similar to that above in Eq.
(\ref{Gaussian}).
 When $\alpha \ll \omega_0$, these sums are dominated by their first
terms, yielding:
\begin{eqnarray}
\sigma^{(s)}_{yy}({\bf k}) =
{\pi \langle |\psi_0|^2 \rangle \over \Phi_0 \Gamma B}
\Biggl[&& 1 +  {3 \hbar^2 \omega_0^2 \ell^2 \over \alpha^2} k_x^2 \nonumber \\
&& -{\ell^2 k_y^2 \over 4} - \ell_c^2 k_z^2  + O(k^4)\Biggr],
\label{sigyy}
\end{eqnarray}
where $\Phi_0=h/2e$ and $\langle |\psi_0|^2 \rangle$, the equilibrium
average of the fluctuations of the order parameter in the LLL, is
\begin{equation}
\langle |\psi_0|^2 \rangle = {k_B T ~m_c^{1/2}
\over 2^{3/2} \pi ~\ell^2 \hbar ~\alpha^{1/2}}.
\label{psi-squared}
\end{equation}

 First note that $S_{yyyy}$ and $S_{yzzy}$ are negative (non-hydrodynamic)
and that the length scales multiplying $k_y^2$ and $k_z^2$ are magnetic
lengths.
 On the contrary, $S_{yxxy}$ is positive and the associated length
scale $\xi_{\perp}\propto \omega_0 \ell/\alpha$ grows as $T$ is lowered.
 So the first candidate for a growing length scale in the resistivity
$\rho_{yy}({\bf r}-{\bf r^{\prime}})$ lies within the $ab$ plane, a
somewhat unexpected result as the picture of moving flux lines
is more suggestive of a long $c$-axis length scale.
 The phase coherence length and the length over which density-density
fluctuations decay have previously been identified as $T$-dependent
length scales in the $ab$ plane\cite{Mike}; recent Monte Carlo simulations
find that they grow in the same way.\cite{Matthew}
 It will prove interesting to follow the renormalization of $\xi_{\perp}$
to determine whether it is distinct from these or not.

 Next note that this high $T$ result for $\sigma^{(s)}_{yy}({\bf 0})$ is
very reminiscent of the flux-flow conductivity $\sigma_{ff}=\eta /\Phi_0 B$
(where $\eta$ is a viscosity coefficient).\cite{Kim}
 The flux-flow conductivity is derived from translating an unpinned
Abrikosov flux-line lattice\cite{Pelcovits}, and its validity depends
on whether one can neglect thermal fluctuations in the vector potential.
 Its similarity to the expression for $\sigma^{(s)}_{yy}$ in eq.
(\ref{sigyy}) occurs since lowering $T$ has minimal effect on
$\sigma^{(s)}_{yy}({\bf 0})$.
 This feature is specific to the $ab$ conductivity; it is not shared
by the $c$-axis conductivity $\sigma^{(s)}_{zz}({\bf 0})$.
 It does extend, however, to the nonlocal conductivity
$\sigma^{(s)}_{yy}( k_x=0,k_y,k_z)$, though it no longer holds
if $k_x \ne 0$.
 We exploit this latter point to argue that $S_{yyyy}$ and $S_{yzzy}$
do not change sign as $T$ is lowered and that the associated length
scales remain short.

So what is this property?
 In the calculation of $\sigma^{(s)}_{yy}({\bf 0})$ the momentum operators
occurring in the definition of the currents (\ref{current}) act like
creation (or annihilation) operators on the $\Psi$'s and $\Psi^*$'s.
 They raise (or lower) the ``states" with the following important
outcome: \ of the two correlators that result, one is in the $n^{th}$
Landau level the other is in the $(n+1)^{th}$.
 In the $\alpha \ll \omega_0$ limit, the contribution with one $n=0$
correlator and one $n=1$ correlator dominates.
(See Fig. 1(a).)

\begin{figure}
\centerline{
\epsfxsize=7cm \leavevmode \epsfbox{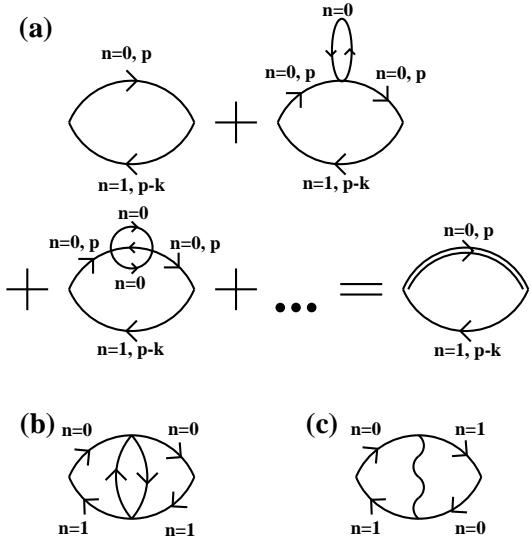}}
\caption{(a) Diagrams which contribute to the $ab$ plane conductivity
in the LLL limit.  (b) Example of a diagram which is ``down.''
(c) Diagram with vector-potential fluctuations which have not been
included.}
\label{fig:diagrams}
\end{figure}

Next recall that the eigenfunction expansion of the correlator
$C({\bf r},{\bf r^{\prime}};t^{\prime}) =
\langle \Psi({\bf r},t+t^{\prime}) \Psi^*({\bf r^{\prime}},t) \rangle$ is
\begin{equation}
C({\bf r},{\bf r^{\prime}};t^{\prime}) \propto
\int_{p_z} \int_{p_y}\sum_n {\phi_n({\bf r})
\phi^*_n({\bf r^{\prime}}) \over E_n(p_z)}
{\rm e}^{-\Gamma E_n(p_z) |t^{\prime}|}.
\label{correlator}
\end{equation}
{}From here we see that in the LLL limit the time integral in the
Kubo formula (\ref{kubo}) becomes
\begin{equation}
\int dt^{\prime} ~{\rm e}^{-\Gamma [E_0(p) + E_1(p-k)]
t^{\prime}},
\label{time-integral}
\end{equation}
where $p$ is the $z$ component of momentum running through the $n=0$
channel and $p-k$ that through the $n=1$ channel.
 (See Fig. 1(a).)
 Note that $E_0$ is much smaller than $E_1$ and we can drop it from
the integral (\ref{time-integral}).
 This last step is equivalent to using the static (equilibrium) $n=0$
correlator and yields the factor of $\langle |\psi_0|^2 \rangle$  in
(\ref{sigyy}).

The main point is that one and only one of the correlators
is in the $n=0$ level and that it can be replaced by a static $n=0$
correlator.
 We now extend these arguments to the nonlocal conductivity.
 Consider $S_{yzzy}$ which featured  in the hydrodynamic
modelling\cite{HusMaj} of the multiterminal transport measurements.
 The only modification in calculating $S_{yzzy}$ is that it requires two
additional derivatives with respect to $z$.
 As these derivatives do not affect the Landau-level structure,
the above arguments carry through.
 In fact, we take further advantage of the disparity in ``masses"
($\alpha \ll \omega_0$) by insisting that the external momentum
$k_z$ be sent through the $n=1$ channel.
 (This choice does not affect the outcome, just makes it more
apparent.)
 The internal momentum integral is dominated by poles associated with
mass $\alpha$, and the terms involving  $k_z$ and $\omega_0$ are left
essentially intact.
 The full $k_z$ dependence in the LLL limit is:
\begin{equation}
\sigma^{(s)}_{yy}( k_z) =
{\pi \langle |\psi_0|^2 \rangle \over \Phi_0 \Gamma B}
\left[1+ k_z^2 \ell_c^2 / 2 \right]^{-2}.
\label{kz-depend}
\end{equation}
The structure is simply that of $[E_1(k_z)]^{-2}$ with $\alpha
\rightarrow 0$.

More detailed calculations reveal that $S_{yyyy}$ has a similar property.
 This time the extra derivatives (with respect to $y$) have some effect
on the Landau-level structure: while one correlator is in the $n=0$ level,
the other is in either the $n=1$ or $n=2$ level.
 The important point is that there is still no contribution with two
$n=0$ correlators.
 This last point applies to all terms in the $k_y$ expansion; the series
resums to give:
\begin{equation}
\sigma^{(s)}_{yy}( k_y) =
{\pi \langle |\psi_0|^2 \rangle \over \Phi_0 \Gamma B}
\left[ {1 - {\rm e}^{ -k_y^2 \ell^2 / 2 }
\over k_y^2 \ell^2 /2} \right].
\label{ky-depend}
\end{equation}
In fact, all the terms in the small-${\bf k}$ expansion of
$\sigma_{yy}(k_y,k_z)$ (so long as $k_x=0$) share this property, and the
combined $k_y$ and $k_z$ dependence is:
\begin{equation}
{\sigma^{(s)}_{yy}(k_y,k_z)
\over \sigma^{(s)}_{yy}({\bf 0})}=
{\rm e}^{- k_y^2 \ell^2 / 2 }
 \sum_{n=0}^{\infty} {(n+1)\left(k_y^2 \ell^2 / 2\right)^n
\over n! \left[ (n+1) + k_z^2 \ell_c^2/ 2 \right]^2}.
\label{k-depend}
\end{equation}
On the other hand, the extra derivatives in $S_{yxxy}$ (with respect
to $x$) have a more crucial effect, producing a nonzero contribution
when both correlators are in the $n=0$ state.
 The outcome, as seen previously, is that $S_{yxxy}$ is positive
(hydrodynamic) and the long length scale $\xi_{\perp}$ arises.

We want to go beyond the Gaussian result, {\it i.e.} consider the effect
of the nonlinear term in the TDGL equation.
 One question is:  To what extent is the simple loop affected?
 In the region of the phase diagram dominated by the LLL, the answer is
``not much," as already hinted at by the similarity of the Gaussian and
flux-flow results.
 Another question is:  Where is this region?
 The Gaussian calculation suggests that the LLL approximation is valid
near the $H_{c2}(T)$ line ($\alpha=0$), but renormalization effects will
change this!
 A very thorough discussion of the region of validity of the LLL appears
in Refs. [7].

Our first attack on the problem of including the nonlinear terms, {\it i.e.}
renormalization effects, will be the Hartree-Fock (HF) approximation.
 The theory remains effectively Gaussian but $a \rightarrow \tilde a =
a +2b\langle |\psi_0^2|\rangle$, giving the following self-consistent
equation for $\tilde \alpha = \tilde a + \hbar \omega_0/2$:
\begin{equation}
\tilde \alpha = \alpha + {b ~k_B T ~m_c^{1/2} \over
2^{1/2} \pi ~\ell^2 \hbar  ~{\tilde \alpha}^{1/2}}.
\label{HF}
\end{equation}
Now we ask not where $\alpha$ is small but where $\tilde \alpha$
is small to establish where the LLL approximation is valid.
 The renormalized mass $\tilde \alpha$ grows small as the bare
mass $\alpha$ grows large and negative, at temperatures below $H_{c2}(T)$.
 Of course, renormalization effects beyond the HF approximation\cite{Ikeda1}
ought to be considered, but concerning the issue of where the
LLL approximation is valid, they are only refinements; the essential
features are captured by HF.

To go beyond the HF approximation to $\sigma^{(s)}_{yy}(k_x=0,k_y,k_z)$,
we expand perturbatively around it.
 Contributions to the $m^{th}$ order require $2m$ additional contractions
($m$ correlators and $m$ propagators).
 The pivotal issue is in what Landau levels do these new correlators and
propagators lie.
 If $\tilde \alpha \ll \omega_0$, then any contribution in which they are in
higher Landau levels would be ``down" (have a significantly lower weight)
compared to those in which they are all in the LLL.
  Any insertion that disrupts the $n \ge 1$ correlator (as in Fig. 1(b))
automatically results in an extra higher Landau-level
contraction.
 The only way in which all new terms are in the LLL is if they only
dress up the $n=0$ correlator.
 This result is shown schematically in Fig. 1(a).

As a consequence, in the LLL approximation, $\sigma_{yy}(k_x=0,k_y,k_z)$
still involves only two correlators: a fully renormalized $n=0$ correlator
$\tilde C^{(0)}({\bf r},{\bf r^{\prime}},t^{\prime})$ and an $n \ge 1$
correlator untouched beyond the HF approximation.
 In the LLL limit $\tilde C^{(0)}({\bf r},{\bf r^{\prime}},t^{\prime})$
has {\it exactly} the same dependence on $x$, $x^{\prime}$, $y$ and
$y^{\prime}$ as its ``bare" version; thus, the $k_y$-dependence
of $\tilde \sigma^{(s)}_{yy}$ is exactly the same as that in
expression (\ref{ky-depend}).
 Moreover, a repetition of the arguments following eq.
(\ref{time-integral}) shows that the $n=0$ correlator can
still be considered static, that the external momentum $k_z$ can
still be sent through the $n=1$ channel and that the internal momentum
integral is still dominated by poles associated with mass $\tilde \alpha$.
 So as before the terms involving the external momentum $k_z$ and mass
$\omega_0$ remain essentially intact.
 The upshot is that the $k_z$-dependence of $\tilde \sigma^{(s)}_{yy}$
is exactly the same as that in expression (\ref{kz-depend}) and
in particular $S_{yzzy}$ and $S_{yyyy}$ remain negative.
 Only $\langle |\psi_0|^2 \rangle$ gets renormalized.

The perturbative arguments above (summarized in Fig. 1(a)) do not hold
when vector-potential fluctuations are no longer negligible.
 Then we would include vertices involving a $\Psi$, a $\Psi^*$ and an
${\cal A}_{\mu}$, where ${\cal A}_{\mu}$ is the fluctuating part of the
vector potential.
 These vertices lead to diagrams like that shown in Fig. 1(c), where
the wavy line represents an ${\cal A}$-${\cal A}$ correlator.
 In the Landau gauge, vertices associated with ${\cal A}_y$ come with a
factor of $\omega_0$ which is large in the LLL limit.
 They also involve a (gauge-invariant) derivative and so the $\Psi$
and $\Psi^*$ at such vertices are in different Landau levels (see Fig.
1(c)).
 Diagrams with ${\cal A}$-${\cal A}$ correlators disrupting the $n=1$
$\Psi$-$\Psi$ correlator (as in Fig. 1(c)) are not down compared to those
with ${\cal A}$-${\cal A}$ correlators dressing up the the $n=0$
correlator.
 Hence the previous arguments no longer apply if vector-potential
fluctuations are important.

The effects of disorder will be considered in a future publication.
 Briefly, point disorder enhances the conductivity in the $ab$ plane;
however, the $c$-axis length scale appears to remain short, $\ell_c$.
 (In a nutshell, the external momentum $k_z$ can still be sent entirely
through an $n=1$ channel in the leading diagrams.)
 On the other hand, in a case with correlated disorder, such as columnar
defects or twin planes, the $c$-axis length scale grows as $T$ is
lowered, suggesting an increased likelihood for nonlocal effects.

In summary the central result is that in the LLL approximation, which
holds over a considerable portion of the $H-T$ diagram for
$YBCO$,\cite{Ikeda1}, the nonlocal conductivity
$\sigma^{(s)}_{yy}(k_y,k_z)$ has the form given in Eq. (\ref{k-depend})
with special cases given by Eqs. (\ref{kz-depend}) and (\ref{ky-depend})
with only $\langle |\psi_0|^2 \rangle$ having any temperature dependence.
 In particular, the length scales multiplying $k_y^2$ and $k_z^2$ in
these expressions are short and temperature-independent; in addition,
the signs of $S_{yyyy}$ and $S_{yzzy}$ remain negative (nonhydrodynamic)
throughout the regime.
 Hence one would expect in experiments on clean samples with geometries
that probe chiefly $\sigma_{yy}$ to see essentially local behavior,
as may be borne out by recent experiments.\cite{Lopez}

\vspace{0.5cm}

We would like to thank the authors of Ref. [13] for fruitful
correspondence and providing their results prior to publication.
 We would also like to thank M.J.W. Dodgson, H. Jensen, A.J. McKane,
N.K. Wilkin and J. Yeo for useful discussions.
 TB acknowledges the support of the Engineering and Physical Science
Research Council under grant GR/K53208.


\bibliographystyle{unsrt}





\end{document}